\documentclass[11pt]{article}
\usepackage{graphics}
\usepackage{epsfig}
\usepackage{citesort}

\usepackage{geometry}
\newcommand{\be}{\begin{eqnarray}}
\newcommand{\ee}{\end{eqnarray}}
\newcommand{\<}{\langle}
\renewcommand{\>}{\rangle}

\newcommand{\nn}{\nonumber}

\begin{document}

\rightline{\Large RM3-TH/05-14}
\rightline{\Large ROME1-1421/2005}

\vspace{1cm}

\begin{center}

\LARGE{\bf Lattice study of semileptonic form factors with twisted boundary conditions}

\vspace{1.5cm}

\Large{D.~Guadagnoli$^a$, F.~Mescia$^{b,c}$ and S.~Simula$^d$}

\vspace{0.5cm}

\normalsize{
$^a$Dipartimento di Fisica, Universit\`a di Roma ``La Sapienza'', 
    and INFN, Sezione di Roma, P.le A.~Moro 2, I-00185 Rome, Italy\\
$^b$Dip. di Fisica, Universit\`a di Roma Tre, Via della Vasca Navale 84, I-00146 Rome, Italy \\
$^c$INFN, Laboratori Nazionali di Frascati, Via E. Fermi 40, I-00044 Frascati, Italy \\
$^d$INFN, Sezione di Roma Tre, Via della Vasca Navale 84, I-00146, Rome, 
Italy}

\end{center}

\vspace{0.5cm}

\begin{abstract}
We apply twisted boundary conditions to lattice QCD simulations of three-point 
correlation functions in order to access spatial components of hadronic momenta 
different from the integer multiples of $2\pi / L$. We calculate the vector and 
scalar form factors relevant to the $K \to \pi$ semileptonic decay and consider all 
the possible ways of twisting one of the quark lines in the three-point 
functions. We show that the momentum shift produced by the twisted boundary 
conditions does not introduce any additional noise and easily allows to determine 
within a few percent statistical accuracy the form factors at quite small values of 
the four-momentum transfer, which are not accessible when periodic boundary conditions 
are considered. The use of twisted boundary conditions turns out to be crucial for a 
precise determination of the form factor at zero-momentum transfer, when a precise 
lattice point sufficiently close to zero-momentum transfer is not accessible with 
periodic boundary conditions.
\end{abstract}

\newpage

\pagestyle{plain}

\section{\protect Introduction\label{sec:introduction}}

In lattice simulations of QCD the spatial components of the hadronic momenta 
$p_j$ ($j = 1, 2, 3$) are quantized. The specific quantized values depend on 
the choice of the boundary conditions (BCs) applied to the quark fields. The 
most common choice is the use of periodic BCs in the spatial directions
 \be 
    \psi(x + \hat{e}_j L) = \psi(x) ~ ,
    \label{eq:periodic}
 \ee
that leads to 
 \be
    p_j = n_j \frac{2 \pi}{L} ~ , 
    \label{eq:pj}
  \ee
where the $n_j$'s are integer numbers. Thus the smallest non-vanishing value 
of $p_j$ is given by $2 \pi / L$, which depends on the spatial size of the 
(cubic) lattice ($V = L^3$). For instance a current available lattice may have 
$L = 32 ~ a$, where $a$ is the lattice spacing, and $a^{-1} \simeq 2.5$ GeV 
leading to $2 \pi / L \simeq 0.5$ GeV. Such a value may represent a strong 
limitation of the kinematical regions accessible for the investigation of 
momentum dependent quantities, like e.g.~form factors.

In Ref.~\cite{Bedaque} it was proposed to use twisted BCs for the quark 
fields
 \be
    \widetilde{\psi}(x + \hat{e}_j L) = e^{2 \pi i \theta_j} ~ \widetilde{\psi}(x)
    \label{eq:twisted}
 \ee
which allows to shift the quantized values of $p_j$ by an arbitrary amount 
equal to $\theta_j 2 \pi/ L$, namely
 \be
    \widetilde{p}_j = \theta_j \frac{2 \pi}{L} + p_j = \theta_j \frac{2 \pi}{L} + n_j 
    \frac{2 \pi}{L} ~ .
    \label{eq:pj_twisted}
  \ee

The twisted BCs (\ref{eq:twisted}) can be shown \cite{Bedaque} to be equivalent 
to the introduction of a U(1) background gauge field coupled to the baryon 
number and applied to quark fields satisfying usual periodic BCs (the Aharonov-Bohm 
effect). Moreover in Ref.~\cite{Tantalo} the twisted BCs were firstly 
implemented in a lattice QCD simulation of two-point correlation functions 
of pseudoscalar mesons. The energy-momentum dispersion relation was checked 
showing that the momentum shift $2 \pi \theta_j / L$ is a true physical one.

The aim of the present paper is to explore the use of distinct twisted 
BCs for different fermion species in the calculation of the momentum 
dependence of lattice three-point correlators\footnote{In this case the 
equivalent background gauge fields are coupled to the generators in the Cartan 
subalgebra of the flavor group $U(N)_V$ commuting with the quark mass matrix.}. 
We want both to establish whether twisted BCs are able to provide in practice 
form factors at small values of the momentum transfer not accessible with periodic 
BCs, and to estimate the level of statistical precision that can be achieved. The 
latter is an important point because the introduction of twisted BCs leads to a 
non-negligible increase of the computational time due to the need of producing 
new inversions of the Dirac equation for each quark momentum. We anticipate that the 
answer to both questions is positive: the twisted BCs are able to provide form factors 
at small values of the momentum transfer with a precision comparable to the one attainable 
with periodic BCs.

In this work we consider the case of the vector and scalar form factors relevant to the 
$K \to \pi$ semileptonic transition, which have been recently investigated using periodic 
BCs by the quenched simulations of Ref.~\cite{Kl3}, as well as by the $n_f = 2$ and 
$n_f = 2 + 1$ dynamical flavor simulations of Refs.~\cite{Kl3_2df} and \cite{Kl3_MILC}, 
respectively. There are two reasons for our choice. 

First, the theoretical uncertainty in the determination of the vector form factor 
at zero-momentum transfer, $f(0)$, presently dominates the corresponding uncertainty in 
the extraction of the value of the Cabibbo angle from $K_{\ell 3}$ decays. As shown 
in Ref.~\cite{Kl3}, an important source of uncertainty for $f(0)$ comes from the 
precision in  the determination of the form factor slopes at zero-momentum transfer. 
Thus an interesting issue is to check whether the access to new lattice points at small 
values of the momentum transfer can improve significantly the precision of the 
determination of $f(0)$.

Second, the approach of Ref.~\cite{Kl3} is characterized by the use of a suitable 
double ratio of three-point correlators, which allows to access in a very precise way the 
scalar form factor $f_0(q^2)$ at $\vec{q} = 0$ corresponding to $q^2 = q_{\rm max}^2 = 
(M_K - M_\pi)^2$. The precision level of the values of $f_0(q_{\rm max}^2)$ is crucial 
to achieve the percent accuracy for $f(0)$. However, there are cases in which it is 
not possible to get a lattice point sufficiently close to $q^2 = 0$ using periodic BCs. 
An important example is represented by some of the form factors entering hyperon 
semileptonic decays, like the weak magnetism, the weak electricity, the induced scalar 
and pseudoscalar form factors \cite{hyperons}. Thus, we have carried out an analysis of 
our lattice data for the $K \to \pi$ transition, but excluding the very accurate value 
$f_0(q_{\rm max}^2)$. We consider this analysis representative of the cases where a 
precise lattice point close to or at zero-momentum transfer is not accessible with 
periodic BCs.

We limit ourselves to quenched simulations where the generation of gauge configurations 
is clearly independent of the BCs applied to the quark fields\footnote{It has been 
recently shown \cite{Sachrajda,Bedaque_2} that for many physical processes, including 
semileptonic decays, one can impose twisted BCs on valence quarks and periodic ones 
for sea quarks, eliminating in this way the need for producing new gauge configurations 
for each quark momentum, since finite-volume effects remain exponentially small.}. 
We expect that the ability of twisted BCs to provide form factors at small values of 
the momentum transfer do not depend on the use of the quenched approximation, and 
therefore we are confident that our findings hold as well also in case of partially 
quenched and full QCD simulations.

We obtain the $K \to \pi$ vector and scalar form factors for quite small values of the 
four-momentum transfer, which are not accessible when periodic BCs are considered, without 
introducing any significant additional noise and within a few percent statistical accuracy. 
For completeness we consider all the possible ways of twisting one of the quark lines in 
the three-point functions.

When the precise lattice point for $f_0(q^2)$ at $q^2 = q_{\rm max}^2$ is not included 
in the analysis, the use of twisted BCs is crucial to allow a determination of $f(0)$
at a few percent statistical level.
On the contrary, when the very accurate value $f_0(q_{\rm max}^2)$ is included in the 
analysis, the impact of twisted BCs on the determination of the $K \to \pi$ form factors 
and their slopes at zero momentum transfer turns out to be marginal, while we remind it 
is expensive for the computational time. This result is due to the fact that: ~ i) the 
precision of the lattice points obtained with twisted BCs is not comparable to the 
one that can be achieved through the double ratio method of Ref.~\cite{Kl3} at 
the particular kinematical point $q^2 = q_{\rm max}^2$, which in turn is quite 
close to $q^2 = 0$ in the simulation of Ref.~\cite{Kl3}; ~ ii) the use of twisted BCs 
does not lead to a sufficient improvement of the precision with respect to the 
nearest space-like points obtained with periodic BCs.

The plan of the paper is as follows. In the next Section we briefly discuss the 
implementation of the twisted BCs for the evaluation of all the propagators 
required in this work. In Section~\ref{sec:correlators} we present the calculation 
of two- and three-point correlation functions having twisted quark lines. 
In Section~\ref{sec:results} we show our results for the scalar and vector 
form factors of the $K \to \pi$ transition, while in Section~\ref{sec:slope} 
we investigate the impact of the twisted BCs on the determination of the slopes 
of the vector and scalar form factors at zero momentum transfer. Finally 
Section~\ref{sec:conclusions} is devoted to our conclusions.

\section{Lattice quark propagators with twisted BCs\label{sec:twisted}}

On the lattice, for a given flavor, the quark propagator $S(x, 0) \equiv 
\< \psi(x) ~ \overline{\psi}(0) \>$, where $\< \dots \>$ indicates the average 
over gauge field configurations, satisfies the following equation
 \be
    \sum_y D(x, y) ~ S(y, 0) = \delta_{x,0}
    \label{eq:Snormal}
  \ee
where $D(x, y)$ is the Dirac operator whose explicit form depends on the choice of the 
lattice QCD action. In what follows we work with Clover fermions and therefore $D(x, y)$ 
is given explicitly by
 \be
    D(x, y) & = & \delta_{x,y} \left( a m_0 + 4r \right) - \frac{1}{2} \sum_{\mu} 
    \left\{ \delta_{x, y - a \hat{\mu}} ~ (r - \gamma_\mu) ~ U_\mu^{\phantom\dagger}(x) 
    \right.  \nn \\
    & + & \left. \delta_{x, y + a \hat{\mu}} ~ (r + \gamma_\mu) ~ U_\mu^\dagger(y)
    \right\} - \frac{c_{SW}~r}{32} ~ \sum_{\mu,\nu} ~ P_{\mu\nu}(x) ~ \sigma_{\mu\nu} 
    ~ \delta_{x,y} ~ ,
    \label{eq:Dirac}
 \ee
where $U_\mu(x)$ is the gauge link, $P_{\mu \nu}(x)$ is the (symmetric) plaquette in the 
($\mu, \nu$) plane and we have omitted Dirac and color indices for simplicity.

When the quark field satisfies the twisted BCs (\ref{eq:twisted}), the corresponding quark 
propagator $\widetilde{S}(x, 0) \equiv \< \widetilde{\psi}(x) ~ \overline{\widetilde{\psi}}(0) 
\>$ still satisfies Eq.~(\ref{eq:Snormal}) with the same Dirac operator $D(x, y)$ but with 
different BCs. Following Refs.~\cite{Bedaque,Tantalo} one can redefine the quark field as 
$\psi_{\vec{\theta}}(x) = e^{- 2 \pi i \vec{\theta} \cdot \vec{x} / L } \widetilde{\psi}(x)$ 
in order to work always with periodic BCs on the fields. In such a way the new quark propagator 
$S^{\vec{\theta}}(x, 0) \equiv \< \psi_{\vec{\theta}}(x) ~ \overline{\psi}_{\vec{\theta}}(0) \> $ 
satisfies the following equation 
 \be
    \sum_y D^{\vec{\theta}}(x, y) ~ S^{\vec{\theta}}(y, 0) = \delta_{x,0}
    \label{eq:Stheta}
  \ee
with a modified Dirac operator $D^{\vec{\theta}}(x, y)$ but periodic BCs. The new Dirac operator 
is related to Eq.~(\ref{eq:Dirac}) by simply rephasing the gauge links
 \be
    U_\mu(x) \to U_\mu^{\vec{\theta}}(x) \equiv e^{2 \pi i a \theta_\mu / L} ~ U_\mu(x)
    \label{eq:rephase}
 \ee
with the four-vector $\theta$ given by ($0, \vec{\theta}$). Note that the plaquette 
$P_{\mu \nu}(x)$ is left invariant by the rephasing of the gauge links. 
In terms of $S^{\vec{\theta}}(x, y)$, related to the quark fields $\psi(x)$ with periodic BCs, 
the quark propagator $\widetilde{S}(x, y)$, corresponding to the quark fields 
$\widetilde{\psi}(x)$ with twisted BCs, is simply given by 
 \be
    \widetilde{S}(x, y) = e^{2 \pi i \vec{\theta} \cdot (\vec{x} - \vec{y}) / L} ~ 
    S^{\vec{\theta}}(x, y) ~ . 
    \label{eq:Stilde}
 \ee

\section{Two- and three-point correlation functions with twisted quark lines\label{sec:correlators}}

We are interested in calculating the $K^0 \to \pi^-$ form factors of the weak vector current 
$V_{\mu} = \bar{s} \gamma_{\mu} u$, which are defined through the relation
 \be
    \langle \pi(p^\prime) | V_{\mu} | K(p)  \rangle = 
    f_+(q^2)(p+p^{\prime})_\mu + f_-(q^2)(p-p^{\prime})_\mu ~ , 
    \label{eq:ff}
 \ee
where $q^2 = (p-p^\prime)^2$. As usual, we express $f_-(q^2)$ in terms of the so-called scalar 
form factor
 \be
    f_0 (q^2) = f_+(q^2)  +  \frac{q^2}{M_K^2 - M_\pi^2}  f_-(q^2) 
    \label{eq:scalar}
 \ee
with $f_{0}(0)=f_+(0)$.

From Eq.~(\ref{eq:ff}) the form factors can be expressed as linear combinations of hadronic 
matrix elements of time and spatial components of the weak vector current. The latter can be 
obtained on the lattice by calculating two- and three-point correlation functions
\be
\label{eq:c3pt}
C_\mu^{K \pi} (t_x,t_y,\vec p,\vec{p}^{\,\prime}) & = & \sum_{\vec x, \vec y} 
\langle O_\pi(t_y,\vec y) ~ \widehat{V}_\mu(t_x,\vec x) ~ O_K^\dagger(0) 
\rangle \, 
e^{-i \vec p \cdot \vec x + i \vec{p}^{\,\prime} \cdot (\vec x - \vec y)}\,\,, \\
C^{K(\pi)} (t,\vec p ) & = & \sum_{\vec x} 
\langle O_{K(\pi)}(t,\vec x) ~ O_{K(\pi)}^\dagger(0) 
\rangle \, 
e^{-i \vec p \cdot \vec x}\,\,,
\ee
where $O_\pi^{\dagger} = \bar d \gamma_5 u$, $O_K^{\dagger} = \bar d \gamma_5 s$ are 
the operators interpolating $\pi^-$ and $K^0$ mesons, and $\widehat{V}_{\mu}$ is 
the renormalized lattice vector current
\be
\widehat{V}^\mu = Z_V \left(1 + b_V \frac{a m_s + a m_\ell}{2} \right) 
\left( \bar{s} \gamma^{\mu} u + c_V \partial_\nu ~ \bar{s} \sigma^{\mu \nu} u 
\right) \,,
\label{eq:vtilde}
\ee
where $Z_V$ is the vector renormalization constant, $b_V$ and $c_V$ are $O(a)$-improvement 
coefficients and the subscript $\ell$ refers to the light $u$ (or $d$) quark. In what 
follows we always use degenerate $u$ and $d$ quarks.

Using the completeness relation and taking $t_x$ and $(t_y - t_x)$ large enough, one gets
\be
C_\mu^{K \pi} (t_x,t_y,\vec p,\vec{p}^{\,\prime}) ~ 
 _{\overrightarrow{\stackrel{\mbox{\tiny $t_x \to \infty$}}{\mbox{\tiny $(t_y - t_x) \to \infty$}}}}
~ \frac{\sqrt{Z_K Z_\pi}} {4 E_K E_\pi} 
\langle \pi(p^\prime) | \widehat{V}_\mu
| K(p) \rangle \, e^{- E_K t_x - E_\pi (t_y - t_x)}\,,
\label{eq:c3ptexp}
\ee
\be
C^{K(\pi)} (t,\vec p (\vec{p}^{\,\prime})) 
& _{\overrightarrow{ {\mbox{\tiny $t \to \infty$}} }} & 
\frac{Z_{K(\pi)}}{2 E_{K(\pi)}} 
e^{- E_{K(\pi)} t}\,,
\label{eq:c2ptexp}
\ee
where $E_K = \sqrt{M_K^2 + |\vec p|^2}$, $E_{\pi} = \sqrt{M_{\pi}^2 + 
|\vec{p}^{\,\prime}|^2}$ and $\sqrt{Z_{K(\pi)}} = \langle 0 | O_{K(\pi)}(0) | K(\pi)
\rangle$. Then it follows
\be
\frac{C_\mu^{K \pi} (t_x,t_y,\vec p,\vec{p}^{\,\prime})}{C^K (t_x,\vec p )
~ C^{\pi} (t_y - t_x, \vec{p}^{\,\prime})} \left.
 _{\overrightarrow{\stackrel{\mbox{\tiny $t_x \to \infty$}}{\mbox{\tiny $(t_y - t_x) \to \infty$}}}}
\right. \frac{\langle \pi(p^\prime) | \widehat{V}_\mu | K(p) \rangle }
{\sqrt{Z_K Z_\pi}}\,.
\label{eq:standard}
\ee
Consequently the hadronic matrix elements $\langle \pi(p^\prime) | \widehat{V}_\mu
| K(p) \rangle$ can be obtained from the plateaux of the l.h.s.~of 
Eq.~(\ref{eq:standard}), once $Z_K$ and $Z_\pi$ are separately extracted from 
the large-time behavior of the two-point correlators (\ref{eq:c2ptexp}).

In terms of the strange and light quark propagators $S_s(x, y)$ and $S_{\ell}(x, y)$ the
two-point correlator $C^{K(\pi)} (t,\vec p (\vec{p}^{\,\prime}))$ becomes
 \be
    C^{K(\pi)} (t, \vec p (\vec{p}^{\,\prime})) = \sum_{\vec x} \langle Tr[ S_{\ell}(x, 0) 
    \gamma_5 S_{s(\ell)}(0, x) \gamma_5 ] \rangle e^{-i \vec p (\vec{p}^{\,\prime}) \cdot \vec x}
    \label{eq:C2pt_PS}
 \ee
where $S_{s(\ell)}(0, x) = \gamma_5 S_{s(\ell)}^{\dagger}(x, 0) \gamma_5$. Analogously the 
three-point correlator $C_\mu^{K \pi} (t_x, t_y, \vec p, \vec{p}^{\,\prime})$ becomes
\be
    C_\mu^{K \pi} (t_x, t_y, \vec p, \vec{p}^{\,\prime}) = \sum_{\vec x} \langle 
    Tr[ \Sigma_{\ell}(x, 0; t_y, \vec{p}^{\,\prime}) \gamma_\mu S_s(0, x)] \rangle 
    e^{-i \vec q \cdot \vec x}
    \label{eq:C3pt_PS}
 \ee
where the {\em generalized} propagator $\Sigma(x, 0; t_y, \vec{p}^{\,\prime})$ satisfies 
the equation \cite{S1}
 \be
    \sum_z D(x, z) ~ \Sigma(z, 0; t_y, \vec{p}^{\,\prime}) = \gamma_5 S(x, 0) 
    e^{-i \vec{p}^{\,\prime} \cdot \vec{x}} ~ \delta_{t_x, t_y} ~ .
    \label{eq:Sigma}
 \ee

\begin{figure}[htb]
\centerline{\includegraphics[scale=0.50]{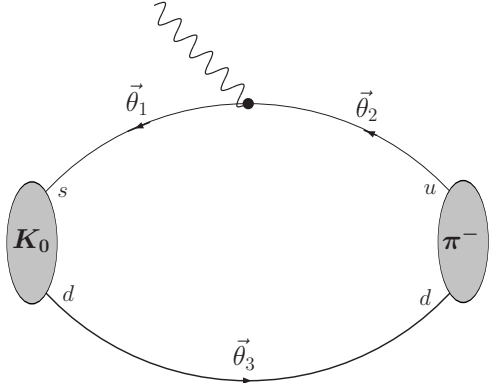}}
\caption{\em Three-point correlation function of the $K^0 \to \pi^-$ semileptonic transition with 
the various quark lines twisted by the vectors $\vec{\theta}_1$, $\vec{\theta}_2$ and $\vec{\theta}_3$.}
\label{fig:theta}
\end{figure}

Using the twisted propagators, $\widetilde{S}$ and $\widetilde{\Sigma}$, 
Eqs.~(\ref{eq:C2pt_PS})-(\ref{eq:Sigma}) hold as well taking into account the corresponding 
change of the quantized momenta ($p_j \to \widetilde{p}_j$). The new two- and three-point 
correlators can be always expressed in terms of quark propagators satisfying periodic BCs, 
namely Eq.~(\ref{eq:Stheta}). For the case of our interest and 
adopting in what follows the convention that the values of the momenta $\vec{p}$ and 
$\vec{p}^{\,\prime}$ are always given by multiples of $2 \pi / L$, we get 
 \be
    C^{K^0}(t, \frac{2 \pi}{L} (\vec{\theta}_3 - \vec{\theta}_1) + \vec p) & = & \sum_{\vec x} \langle 
    Tr[ S_d^{\vec{\theta}_3}(x, 0) \gamma_5 S_s^{\vec{\theta}_1}(0, x) 
    \gamma_5 ] \rangle e^{-i \vec p \cdot \vec x} \nonumber \\
    C^{\pi^-}(t, \frac{2 \pi}{L} (\vec{\theta}_3 - \vec{\theta}_2) + \vec{p}^{\,\prime}) & = & 
    \sum_{\vec x} \langle Tr[ S_d^{\vec{\theta}_3}(x, 0) \gamma_5 S_u^{\vec{\theta}_2}(0, x) 
    \gamma_5 ] \rangle e^{-i \vec{p}^{\,\prime} \cdot \vec x}
    \label{eq:C2pt_twisted}    
 \ee
 \be
    C_\mu^{K^0 \pi^-} (t_x, t_y, \frac{2 \pi}{L} (\vec{\theta}_3 - \vec{\theta}_1) + \vec p, 
    \frac{2 \pi}{L} (\vec{\theta}_3 - \vec{\theta}_2) + \vec{p}^{\,\prime}) = \sum_{\vec x} 
    e^{-i \vec q \cdot \vec x} \nonumber \\
    \langle Tr[ \Sigma_{ud}^{\vec{\theta}_2, \vec{\theta}_3}(x, 0; t_y, \vec{p}^{\,\prime}) 
    \gamma_\mu S_s^{\vec{\theta}_1}(0, x)] \rangle \nonumber \\[4mm]
    C_\mu^{\pi^- K^0 } (t_x, t_y, \frac{2 \pi}{L} (\vec{\theta}_3 - \vec{\theta}_1) + \vec p, 
    \frac{2 \pi}{L} (\vec{\theta}_3 - \vec{\theta}_2) + \vec{p}^{\,\prime}) = \sum_{\vec x} 
    e^{-i \vec q \cdot \vec x} \nonumber \\
    \langle Tr[ \Sigma_{sd}^{\vec{\theta}_2, \vec{\theta}_3}(x, 0; t_y, \vec{p}^{\,\prime}) 
    \gamma_\mu S_u^{\vec{\theta}_1}(0, x)] \rangle 
    \label{eq:C3pt_twisted}
 \ee
where $\vec{q} = \vec{p} - \vec{p}^{\,\prime}$ and the {\em generalized} propagator 
$\Sigma_{q_2 q_3}^{\vec{\theta}_2, \vec{\theta}_3}$ is solution of the modified equation
 \be
    \sum_z D_{q_2}^{\vec{\theta}_2}(x, z) ~ \Sigma_{q_2 q_3}^{\vec{\theta}_2, \vec{\theta}_3}(z, 0; t_y, 
    \vec{p}^{\,\prime}) = \gamma_5 S_{q_3}^{\vec{\theta}_3}(x, 0) e^{-i \vec{p}^{\,\prime} \cdot \vec{x}} 
    ~ \delta_{t_x, t_y} ~ .
    \label{eq:Sigma_twisted}
 \ee
To help visualizing the notation used for the $\theta$-vectors the three-point correlator 
relevant for the $K^0 \to \pi^-$ semileptonic transition is depicted in Fig.~\ref{fig:theta} .

Note that in the r.h.s.~of Eqs.~(\ref{eq:C2pt_twisted})-(\ref{eq:Sigma_twisted}) the exponentials 
do not contain any $\theta$-vector because of Eq.~(\ref{eq:Stilde}).

\section{\boldmath Results for the $K \to \pi$ form factors\label{sec:results}}

We have generated $100$ quenched gauge field configurations on a $24^3 \times 56$ lattice at
$\beta = 6.20$ (corresponding to an inverse lattice spacing equal to $a^{-1} \simeq 2.6$ GeV),
with the plaquette gauge action. Using non-perturbatively $O(a)$-improved Wilson fermions we 
have chosen quark masses corresponding to one pair of the values adopted in Ref.~\cite{Kl3},
namely $k \in \{ 0.1339, 0.1349 \}$.

Using $K$ and $\pi$ mesons with quark content ($k_s k_\ell$) and ($k_\ell k_\ell$) 
respectively, two different $K \to \pi$ correlators ($C_\mu^{K \pi}$) have been computed, 
using both $k_s < k_\ell$ and $k_s > k_\ell$, corresponding to the cases in which the 
kaon(pion) is heavier than the pion(kaon). Using the same combinations of quark masses, 
also the three-point $\pi \to K$ correlations ($C_\mu^{\pi K}$) have been calculated. 
Finally, two non-degenerate $K \to K$ and two degenerate $\pi \to \pi$ three-point 
functions have been evaluated.

As for the critical hopping parameter, we have adopted the value $k_c = 0.135820(2)$ found in 
Ref.~\cite{Kl3} using the axial Ward identity, and we have chosen the time insertion of 
the vector current equal to $t_y = T/2$, which allows to average the three-point correlators 
between the left and right halves of the lattice. Using the degenerate $K \to K$ and $\pi \to \pi$ 
transitions, for which the vector form factor at zero-momentum transfer is known to be equal to unity 
because of vector current conservation, the values of the renormalization constant $Z_V$ and of 
the $O(a)$-improvement parameter $b_V$, appearing in Eq.~(\ref{eq:vtilde}), are found to be in good 
agreement with the corresponding results of Ref.~\cite{Kl3}. Finally, we adopt for the improvement 
coefficient $c_V$ the same non-perturbative value $c_V = -0.09$ used in Ref.~\cite{Kl3}. Throughout 
this paper the statistical errors are evaluated using the jacknife procedure.

As shown in Ref.~\cite{Kl3} the scalar form factor $f_0(q^2)$ can be calculated very efficiently 
at $q^2 = q_{\rm max}^2 = (M_K - M_\pi)^2$ using a double ratio of three-point correlation 
functions with both mesons at rest, namely
\be
   R_0(t_x, t_y) \equiv \frac{C_0^{K \pi}(t_x,t_y,\vec 0,\vec 0) \, C_0^{\pi K}
   (t_x,t_y,\vec 0,\vec 0)} {C_0^{K K}(t_x,t_y,\vec 0,\vec 0) \, C_0^{\pi \pi}
   (t_x,t_y,\vec 0,\vec 0)} ~ 
   _{\overrightarrow{\stackrel{\mbox{\tiny $t_x \to \infty$}}{\mbox{\tiny $(t_y - t_x) \to \infty$}}}}  
   ~ [f_0(q^2_{\rm max})]^2 \,\frac{(M_K + M_\pi)^2}{4 M_K M_\pi} ~ .
   \label{eq:fnal}
\ee

For $q^2 \neq q_{\rm max}^2$ the form factors $f_+(q^2)$ and $f_0(q^2)$ can be determined from 
the matrix elements of the time and spatial components of the vector current, $\langle \pi | 
\widehat{V}_0 | K \rangle$ and $\langle \pi | \widehat{V}_i | K \rangle$ ($i = x, y, z$), that 
in turn can be extracted from the plateaux of the l.h.s.~of Eq.~(\ref{eq:standard}) (as discussed 
in the previous Section). However, such a plain strategy leads to a determination of $f_0(q^2)$ with 
a quite poor precision (see Ref.~\cite{Kl3}). Therefore in order to achieve a much better 
accuracy we follow Ref.~\cite{Kl3} and introduce a suitable ratio of the spatial and time components 
$\langle \pi | \widehat{V}_0 | K \rangle$ and $\langle \pi | \widehat{V}_i | K \rangle$, normalized by the corresponding 
degenerate $K \to K$ transition, namely
 \be
    R_{i0} \equiv \frac{\langle \pi | \widehat{V}_i | K \rangle}{\langle \pi | \widehat{V}_0 
    | K \rangle} \frac{\langle K | \widehat{V}_0 | K \rangle}{\langle K | \widehat{V}_i | 
    K \rangle} ~ ,
    \label{eq:Rio}
 \ee
which allows to access the ratio $f_-(q^2) / f_+(q^2)$.

First of all, as a consistency check of our run, we have evaluated the two- and three-point 
correlation functions (\ref{eq:C2pt_PS}) and (\ref{eq:C3pt_PS}) setting all the three 
$\theta$-vectors to zero and adopting the same values of $\vec{p}$ and $\vec{p}^{\,\prime}$ 
used in Ref.~\cite{Kl3}, where a different, larger set of $230$ quenched gauge configurations 
were employed. The results obtained for both $f_0(q_{\rm max}^2)$ and the scalar and vector 
form factors at $q^2 \neq q_{\rm max}^2$ are nicely consistent within the statistical errors. 
In what follows we will label these results as ``$\theta = 0$''. Then, the two- and three-point 
correlation functions (\ref{eq:C2pt_twisted}) and (\ref{eq:C3pt_twisted}) have been evaluated 
choosing always $\vec{p} = \vec{p}^{\,\prime} = 0$ and making three different kinematical choices 
that correspond to assuming non-vanishing only one out of the three $\theta$-vectors of 
Fig.~\ref{fig:theta}. Defining $Q \equiv  2 \pi |\vec{\theta}| / L$, $E_K \equiv \sqrt{M_K^2 + 
Q^2}$ and $E_\pi \equiv \sqrt{M_\pi^2 + Q^2}$ we consider

\begin{itemize}

\item {\bf Kinematics A:} $\vec{\theta}_1 = \vec{\theta} \neq 0$ and $\vec{\theta}_2 = 
\vec{\theta}_3 = 0$ ($\Rightarrow \vec{p}_K =  2 \pi \vec{\theta} / L$, $\vec{p}_{\pi} = \vec{0}$)
 \be
    f_+(q^2) & = & \frac{1}{2M_\pi} \left[ \langle \pi | \widehat{V}_0 | K \rangle - 
    \frac{E_K - M_\pi}{Q} \langle \pi | \widehat{V}_i | K \rangle \right] \nonumber \\[2mm]
    f_0(q^2) & = & f_+(q^2) \left[ 1 + \frac{q^2}{M_K^2 - M_\pi^2} \frac{(E_K + M_\pi) R_{i0} -
    (M_K + E_K)}{M_K + E_K - (E_K - M_\pi) R_{i0}} \right] \nonumber \\[2mm]
    q^2 & = & M_K^2 + M_\pi^2 - 2 M_\pi E_K = q_{\rm max}^2 - 2 M_\pi ( E_K - M_K) ~ ;
    \label{eq:ffA}
 \ee

\item {\bf Kinematics B:} $\vec{\theta}_2 = \vec{\theta} \neq 0$ and $\vec{\theta}_1 = 
\vec{\theta}_3 = 0$ ($\Rightarrow \vec{p}_K = \vec{0}$, $\vec{p}_{\pi} =  2 \pi \vec{\theta} / L$)
 \be
    f_+(q^2) & = & \frac{1}{2M_K} \left[ \langle \pi | \widehat{V}_0 | K \rangle - 
    \frac{E_\pi- M_K}{Q} \langle \pi | \widehat{V}_i | K \rangle \right] \nonumber \\[2mm]
    f_0(q^2) & = & f_+(q^2) \left[ 1 + \frac{q^2}{M_K^2 - M_\pi^2} \frac{M_K + E_K - 
    (E_\pi + M_K) R_{i0}}{M_K + E_K - (E_\pi - M_K) R_{i0}} \right] \nonumber \\[2mm]
    q^2 & = & M_K^2 + M_\pi^2 - 2 M_K E_\pi = q_{\rm max}^2 - 2 M_K ( E_\pi - M_\pi) ~ ;
    \label{eq:ffB}
 \ee

\item {\bf Kinematics C:} $\vec{\theta}_3 = \vec{\theta} \neq 0$ and $\vec{\theta}_1 = 
\vec{\theta}_2 = 0$ ($\Rightarrow \vec{p}_K = \vec{p}_{\pi} =  2 \pi \vec{\theta} / L$)

In this case the vector form factor is given by $f_+(q^2) = \langle \pi | \widehat{V}_i | K \rangle / 2Q$; 
however a more accurate determination can be obtained by constructing a double ratio similar 
to the one in Eq.~(\ref{eq:fnal}), but using the spatial components of the weak vector current, 
namely
 \be
   R_i(t_x, t_y) \equiv \frac{C_i^{K \pi}(t_x,t_y,\vec{p}_K,\vec{p}_{\pi}) \, C_i^{\pi K}
   (t_x,t_y,\vec{p}_{\pi},\vec{p}_K)} {C_i^{K K}(t_x,t_y,\vec{p}_K,\vec{p}_K) \, C_i^{\pi \pi}
   (t_x,t_y,\vec{p}_{\pi},\vec{p}_{\pi})} ~ 
   _{\overrightarrow{\stackrel{\mbox{\tiny $t_x \to \infty$}}{\mbox{\tiny $(t_y - t_x) \to \infty$}}}}  
   ~ [f_+(q^2)]^2
   \label{eq:fnal_i}
 \ee
 and
 \be
    f_0(q^2) & = & f_+(q^2) \left[ 1 + \frac{q^2}{M_K^2 - M_\pi^2} \frac{1}{E_K - E_\pi} 
    \left(\frac{2 E_K}{R_{i0}} - E_K - E_\pi \right) \right] \nonumber \\[2mm]
    q^2 & = & (E_K - E_\pi)^2 ~ .
    \label{eq:ffC}
 \ee

\end{itemize}

Notice that in kinematics C, by varying $|\vec{\theta}|$, the values of $q^2$ are always 
positive in the range $0 \div q_{\rm max}^2$.

We consider two values of $|\vec{\theta}|$, namely $|\vec{\theta}| = 0.225, 0.70$. The 
first value leads to a difference $\sqrt{M_{K(\pi)}^2 + ( 2 \pi |\vec{\theta}| / L)^2} - 
M_{K(\pi)}$ just exceeding the statistical errors, while the second one simply gives rise to 
a value of $q^2$ which is approximately half of the minimum, non-vanishing value attainable 
for space-like $q^2$ with periodic BCs.

For each of the above values we consider two different orientations: $\vec{\theta} = (1, 0, 0) 
\cdot |\vec{\theta}|$ (asymmetric) and $\vec{\theta} = (1, 1, 1) \cdot |\vec{\theta}| / \sqrt{3}$ 
(symmetric). In this way we can check whether either spatially asymmetric or symmetric momentum 
shifts lead to different noises. We have found no significant difference.

In Table~\ref{tab:kin} we have collected the values of the meson masses, the SU(3)-breaking 
parameter ($M_K^2 - M_\pi^2$) and of $q_{\rm max}^2$ in lattice units that characterize our 
simulation. We remind that for each value of $\vec{\theta}$ a new inversion of Eq.~(\ref{eq:Stheta}) 
is required with a computational time similar to the one needed for the inversion of Eq.~(\ref{eq:Snormal}) 
at $\vec{\theta} = 0$.

\begin{table}[htb]

\begin{center}
\vspace{0.25cm}
\begin{tabular}{||c||c|c||c|c||}
\hline 
$k_s - k_\ell$ & $a M_K$ & $a M_\pi$ & $a^2 (M_K^2 - M_\pi^2)$ & $a^2 q_{\rm max}^2$ \\ \hline
 $0.1339-0.1349$ & $0.3025(20)$ & $0.2419(24)$ & $+0.03299(26)$ & $0.00367(8) $\\ \hline
 $0.1349-0.1339$ & $0.3025(20)$ & $0.3556(15)$ & $-0.03495(30)$ & $0.00282(8) $\\ \hline
\end{tabular}
\end{center}
 
\caption{\it Values of the hopping parameters $k_s$ and $k_\ell$, $a^2 (M_K^2 - M_\pi^2)$,
and $a^2 q_{\rm max}^2$.}
 
\label{tab:kin}
 
\end{table}

\begin{figure}[htb]
\includegraphics[bb=-1cm 17cm 32cm 29cm, scale=0.650]{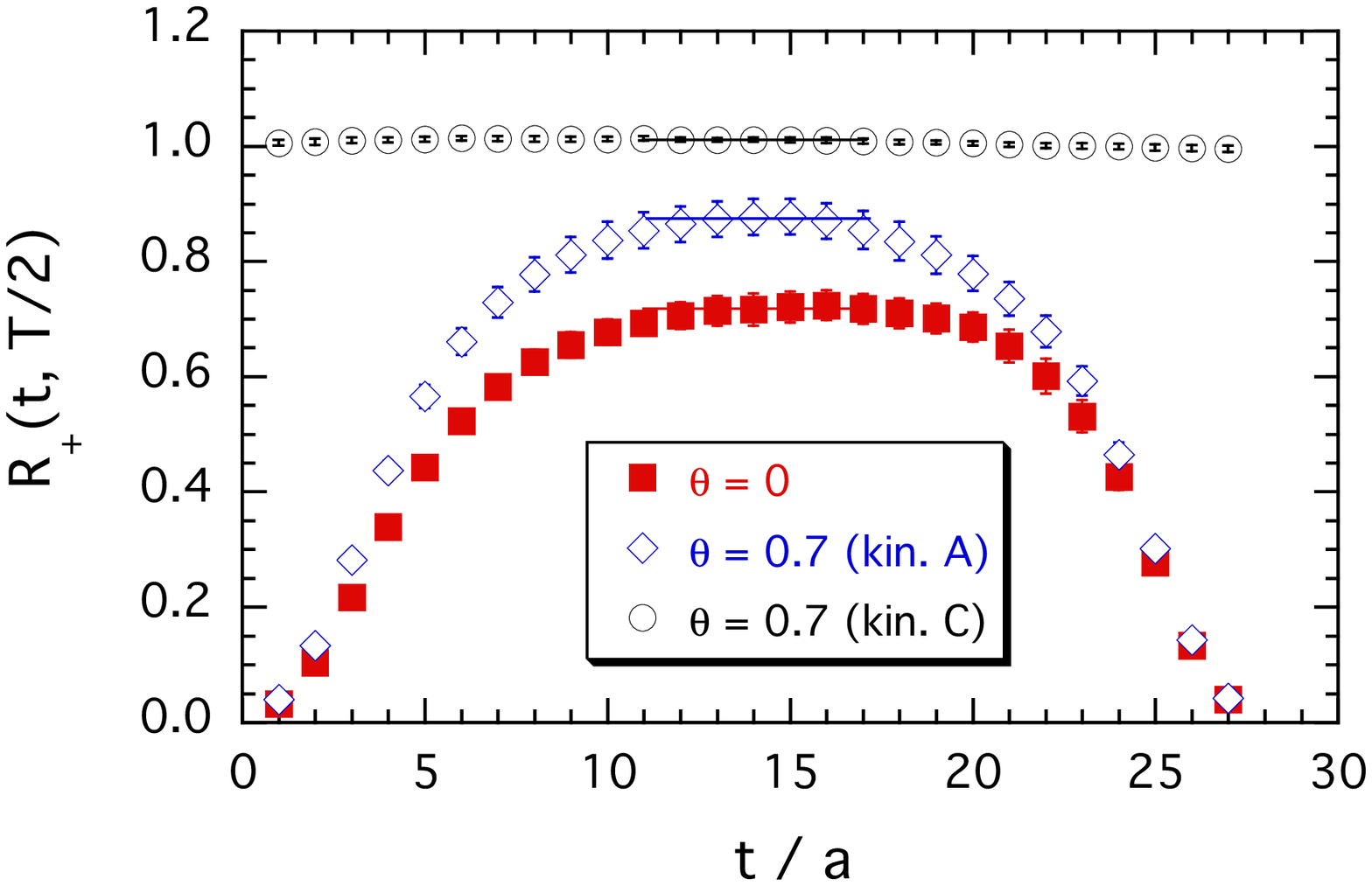}
\caption{\em Time dependence of the quantity $R_+(t_x = t, t_y = T/2)$ providing at large 
time distances the vector form factor $f_+(q^2)$ (see footnote~\ref{fn:R+} in the 
text), calculated with periodic BCs ($\theta = 0$) at $a^2 q^2 = - 0.067$ (full squares), 
and with twisted BCs ($\vec{\theta} = (0.7, 0, 0)$) at $a^2 q^2 = - 0.034$ in kinematics A 
(open diamonds) and $a^2 q^2 = 0.0022$ in kinematics C (open dots). The plateaux are taken 
from $t / a = 11$ to $t / a = 17$. The values of the hopping parameters are $k_s = 
0.1349$ and $k_{\ell} = 0.1339$.}
\label{fig:plateaux}
\end{figure}

The quality of the plateaux used for extracting the vector form factor $f_+(q^2)$ can be 
appreciated in Fig.~\ref{fig:plateaux} for three representative cases corresponding to 
$\theta = 0$ and $\vec{\theta} = (0.7, 0, 0)$ in kinematics A and C\footnote{\label{fn:R+} 
For sake of simplicity we do not report explicitly the definition of the quantity 
$R_+(t_x, t_y)$ appearing in Fig.~\ref{fig:plateaux} for each kinematics. It suffices to 
say that it is defined in terms of the l.h.s.~of Eq.~(\ref{eq:standard}) for the various 
components of the weak current which should be combined to determine the vector form factor 
in the various kinematics [see Eqs.~(\ref{eq:ffA}-\ref{eq:fnal_i})]. In kinematics C 
$R_+(t_x, t_y)$ is defined as $\sqrt{R_i(t_x, t_y)}$ [see Eq.~(\ref{eq:fnal_i})].}. Note 
that in the latter kinematics he use of the double ratio (\ref{eq:fnal_i}) allows to 
reduce strongly statistical fluctuations.

\begin{figure}[htb]
\includegraphics[bb=1cm 20cm 33cm 30cm, scale=0.80]{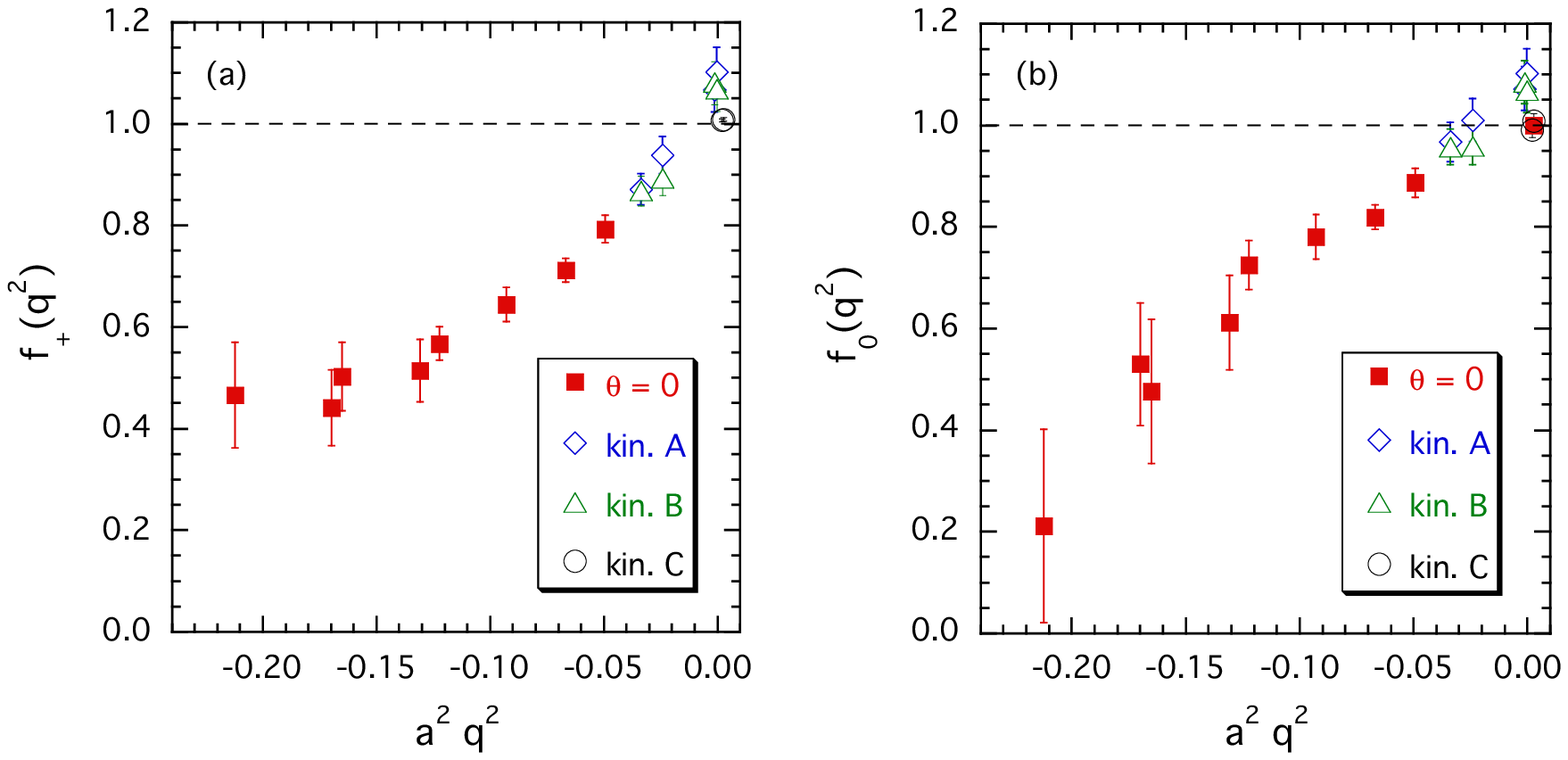}
\caption{\em Results for the form factors $f_+(q^2)$ (a) and $f_0(q^2)$ (b) obtained for various 
values of $q^2$ at $k_s = 0.1349$ and $k_\ell = 0.1339$. Full squares correspond to $\theta = 0$, 
while open diamonds, triangles and dots correspond to the three kinematics A, B and C, respectively.}
\label{fig:ffs}
\end{figure}

\begin{figure}[htb]
\includegraphics[bb=1cm 20cm 33cm 30cm, scale=0.80]{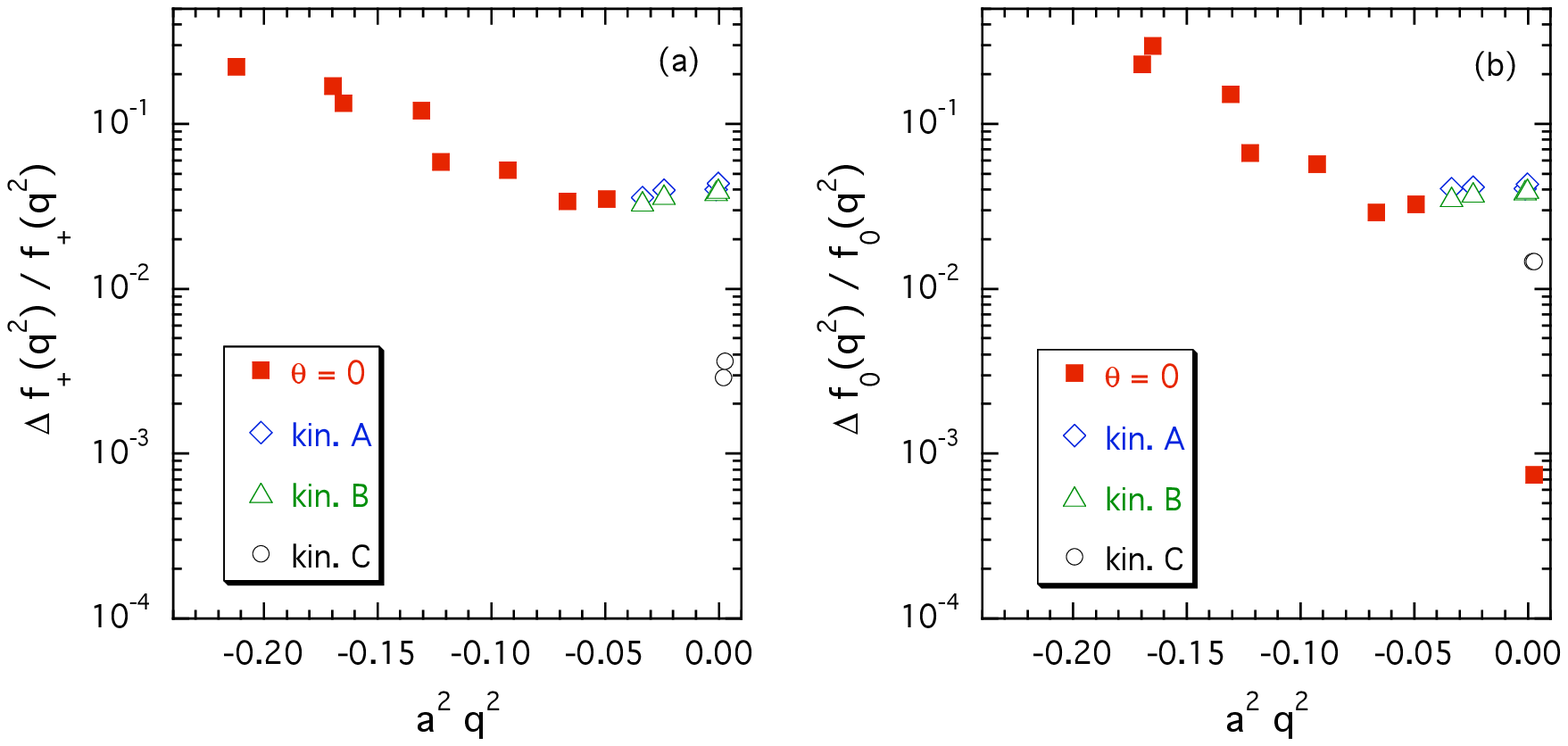}
\caption{\em As in Fig.~\ref{fig:ffs} but for the relative statistical errors $\Delta f_+(q^2) / 
f_+(q^2)$ (a) and $\Delta f_0(q^2) / f_0(q^2)$ (b).}
\label{fig:ffs_delta}
\end{figure}

In Fig.~\ref{fig:ffs} our results for the form factors $f_+(q^2)$ and $f_0(q^2)$, obtained 
at $k_s = 0.1349$ and $k_\ell = 0.1339$ for the three kinematics A, B and C as well as 
for $\theta = 0$, are reported, while in Fig.~\ref{fig:ffs_delta} the relative statistical 
errors, $\Delta f_+(q^2) /f_+(q^2)$ and $\Delta f_0(q^2) /f_0(q^2)$, are shown. It can be seen 
that the use of twisted BCs allows to explore the low-$q^2$ region without introducing any 
significant additional noise. 

The statistical error of the results obtained at $\theta = 0$ quickly decreases as $q^2$ 
increases in the space-like region, reaching a minimum value of $\simeq 3 \%$ at $a^2 q^2 
\simeq = -0.05$ for both $f_+(q^2)$ and $f_0(q^2)$. 
The results obtained using kinematics A and B are totally consistent with each other and 
the statistical error remains almost constant ($\simeq 4 \%$). Note that 
the precision obtained in kinematics A and B is not better than the one 
of the nearest points obtained at $\theta = 0$. 

Vice versa a significant improvement in the precision can be achieved using 
kinematics C thanks to the double ratio (\ref{eq:fnal_i}), which allows to 
determine $f_+(q^2)$ with a statistical error of $\simeq 0.3 \%$. For the 
scalar form factor the corresponding accuracy is only $\simeq 1.5 \%$ due 
to the larger fluctuations of the ratio (\ref{eq:Rio}), but it is still 
better than the precision achieved with kinematics A and B.
The smallest statistical uncertainty ($\simeq 0.07 \%$) remains the one at 
$q^2 = q_{\rm max}^2$ thanks to the double ratio (\ref{eq:fnal}) that involves 
the time component of the weak vector current with all mesons {\em at rest}.

\section{\boldmath Slopes of $f_+(q^2)$ and $f_0(q^2)$ at $q^2 = 0$\label{sec:slope}}

In this Section we analyze the momentum dependence of the vector and scalar $K \to \pi$ 
form factors in order to understand the impact of the introduction of the twisted BCs in
the determination of the form factors as well as of their slopes at 
zero-momentum transfer. As already noted in Ref.~\cite{Kl3} the results for $f_+(q^2)$ 
can be very well described by a pole-dominance fit, where the slope agrees well with 
the inverse of the $K^*$-meson mass square for each combination of the simulated quark 
masses. The results for $f_0(q^2)$ can be parameterized using different functional forms, 
like the ones adopted in Ref.~\cite{Kl3}. However, for our purposes it suffices to consider 
the case of a polar fit for both $f_+(q^2)$ and $f_0(q^2)$. We have checked that 
our final findings remain unchanged if instead of a polar fit  a linear or quadratic 
$q^2$-dependence for $f_0(q^2)$ is considered.

Thus our results have been parameterized using the following momentum dependencies
 \be
    f_+(q^2) & = & \frac{f(0)}{1 - \lambda_+ ~ q^2} ~ , \nonumber \\[2mm]
    f_0(q^2) & = & \frac{f(0)}{1 - \lambda_0 ~ q^2} ~ , 
    \label{eq:polar}
 \ee
where $f(0) = f_+(0) = f_0(0)$, $\lambda_0$ and $\lambda_+$ are fitting parameters.
We have determined the values of such three parameters through a $\chi^2$-minimization 
procedure applied to different sets of lattice data, namely: i) data obtained using 
periodic BCs only ($\theta = 0$); ii) addition of the results corresponding to kinematics 
A and B; iii) data at $\theta = 0$ plus those corresponding to kinematics C only; iv) 
full set of lattice points ($\theta = 0$ plus kinematics A, B and C). In Table~\ref{tab:slopes} 
we have reported the values obtained for $f(0)$, $\lambda_0$ and $\lambda_+$ for each choice 
of the data set. It can be seen that the impact of the lattice points corresponding to 
twisted BCs appears to be marginal and the extracted values of $f(0)$, $\lambda_0$ and 
$\lambda_+$ have almost the same accuracy as the one obtained using simply the lattice 
data calculated with periodic BCs. The usefulness of the twisted BCs is thus spoiled by 
the absence of a significant improvement of the accuracy as well as by the non-negligible 
increase of the computational time required for the inversions of Eq.~(\ref{eq:Stheta}) for 
each choice of $\vec{\theta}$.

\begin{table}[htb]
\begin{center}
\begin{tabular}{||c||c||c|c||} \hline 
{\rm lattice data set} & $f(0)$ & $\lambda_0 / a^2$ & $\lambda_+ / a^2$ \\ \hline \hline
$\theta = 0$ & 0.9907(19) & 3.2(5) & 5.9(8) \\ \hline \hline
$\theta = 0 ~ + $ kin.~A, B & 0.9920(17) & 2.8(5) & 5.6(7) \\ \hline \hline
$\theta = 0 ~ + $ kin.~C & 0.9910(16) & 3.1(5) & 6.0(8) \\ \hline \hline
$\theta = 0 ~ + $ kin.~A, B, C & 0.9921(15) & 2.7(4) & 5.6(7) \\ \hline \hline
\end{tabular} 
\end{center}

\caption{\em Values of the vector form factor at zero-momentum transfer $f(0) = f_+(0) = f_0(0)$ 
and of the slope parameters $\lambda_0$ and $\lambda_+$ of the polar fits (\ref{eq:polar}), 
obtained using different sets of values of $f_+(q^2)$ and $f_0(q^2)$, including always the 
very accurate value of $f_0(q^2)$ at $q^2 = q_{\rm max}^2$. The values of the hopping 
parameters are $k_s = 0.1349$ and $k_{\ell} = 0.1339$.}

\label{tab:slopes}

\end{table}

This finding can be traced back to the following facts: 
i) the presence of the very accurate value of $f_0(q^2)$ at $q^2 = 
q_{\rm max}^2$ in all the data sets considered (such a data point is 
by far the most accurate one [see the rightmost point in Fig.~\ref{fig:ffs_delta}(b)] 
and also the values of $q_{\rm max}^2$ are quite small [see Table~\ref{tab:kin}]); 
and ii) the use of twisted BCs does not lead to a sufficient improvement 
of the precision with respect to the nearest space-like points obtained with 
periodic BCs.

As mentioned in the Introduction, there are cases of phenomenological interest where a 
lattice point at zero-momentum transfer is not accessible with periodic BCs. An important 
example is represented by some of the form factors entering hyperon semileptonic decays, 
like the weak magnetism, the weak electricity, the induced scalar and pseudoscalar form 
factors \cite{hyperons}. Thus, in order to clarify the role played by the $\vec{\theta} 
\neq 0$ lattice points, we have repeated the fitting procedure for the same lattice data 
sets but excluding always the very accurate value $f_0(q_{\rm max}^2)$. This analysis is 
expected to be representative of the cases where a precise lattice point close to or at 
zero-momentum transfer is not accessible. Our results are reported in 
Table~\ref{tab:slopes_noqmax}.

\begin{table}[htb]
\begin{center}
\begin{tabular}{||c||c||c|c||} \hline 
{\rm lattice data set} & $f(0)$ & $\lambda_0 / a^2$ & $\lambda_+ / a^2$ \\ \hline \hline
$\theta = 0$ & 1.107(105) & 5.2(2.1) & 8.1(2.4) \\ \hline \hline
$\theta = 0 ~ + $ kin.~A, B & 1.089~(47) & 4.7(8) & 7.7(9) \\ \hline \hline
$\theta = 0 ~ + $ kin.~C & 0.9934(23) & 3.3(6) & 6.0(8) \\ \hline \hline
$\theta = 0 ~ + $ kin.~A, B, C & 0.9966(25) & 3.0(5) & 5.5(7) \\ \hline \hline
\end{tabular}
\end{center}

\caption{\em The same as in Table~\ref{tab:slopes} but excluding in the data sets 
the value of $f_0(q^2)$ at $q^2 = q_{\rm max}^2$.}

\label{tab:slopes_noqmax}

\end{table}

 The following comments are in order:
\begin{itemize}

\item using the data set $\theta = 0$ (only periodic BCs) the values of $f(0)$ and of the 
slopes are now more poorly determined. The statistical uncertainties turn out to be $\simeq 10 \%$, 
$\simeq 40 \%$ and $\simeq 30 \%$ for $f(0)$, $\lambda_0$ and $\lambda_+$, respectively. Note 
that the accuracy obtained for $f(0)$ is almost $3$ times the precision of the nearest space-like 
points ($\simeq 3 \%$);

\item the accuracy improves by a factor of $\simeq 2$ when the data corresponding to kinematics 
A and B are included. Note that the accuracy of $f(0)$ is now comparable to the one of the 
$\vec{\theta} \neq 0$ points ($\simeq 4 \%$), due to the fact that the latter are quite close 
to $q^2 = 0$. Clearly the $\simeq 5 \%$ precision achieved for $f(0)$ is not enough for 
phenomenological applications to the $K_{\ell 3}$ decay. However such a level of accuracy 
is highly desirable for other observables, like hyperon semileptonic form factors;

\item the inclusion of the quite accurate points obtained in kinematics C (with or without those 
of kinematics A and B) leads to a remarkably good determination of $f(0)$ and the slopes $\lambda_0$ 
and $\lambda_+$, obtaining a precision competitive with that reported in Table~\ref{tab:slopes}, 
where the very precise value $f_0(q_{\rm max}^2)$ is included.

\end{itemize}

We have also performed the analysis of our lattice simulations at $k_s = 0.1339$ and 
$k_{\ell} = 0.1349$. The results obtained are similar to the ones shown in 
Figs.~\ref{fig:plateaux}-\ref{fig:ffs_delta} and in Tables~\ref{tab:slopes}-\ref{tab:slopes_noqmax}. 
Our findings about the impact of the twisted BCs remain unchanged. The same is true also for 
the $K \to K$ and $\pi \to \pi$ transitions, where $q_{\rm max}^2$ vanishes, 
$f_-(q^2) = 0$ and $f(0) = 1$ because of vector current conservation. Note that for degenerate 
transitions, like the ones needed for the nucleon magnetic and the neutron electric dipole 
form factors, the kinematics A and B precisely coincide, while the kinematics C does not add any 
new information.

We mention that there are other cases of phenomenological interest where a lattice 
point at zero-momentum transfer is not accessible with periodic BCs. For instance the 
nucleon magnetic form factor $G_M(q^2)$ is related to the matrix elements of the 
spatial components of the electromagnetic current, which are proportional to the value 
of the momentum transfer. Therefore, the nucleon magnetic moment $G_M(q^2 = 0)$ is not 
directly accessible with periodic BCs and it should be determined by a ``long" 
extrapolation to $q^2 = 0$ (see Ref.~\cite{nucleon}). Another example is the neutron 
electric dipole moment as determined with the strategies described in 
Refs.~\cite{NEDM_1,NEDM_2}, which require again a long extrapolation of the 
$CP$-violating neutron form factor $F_3(q^2)$ to $q^2 = 0$~\footnote{A direct 
determination of the form factor $F_3(q^2)$ at $q^2 = 0$ can be obtained with the 
approach proposed in Ref.~\cite{NEDM_3}.}. In such cases however the introduction of 
twisted BCs is not trivial, because the current involved, the electromagnetic one, is 
not flavor changing. One can speculate that, by introducing an additional flavor and 
suitable interpolating fields, the application of the twisted BCs to the additional 
flavor, at least for quenched simulations, might provide the form factor of interest 
at quite small values of the four-momentum transfer. The application of twisted BCs 
to electromagnetic transitions requires therefore a careful treatment, which is well 
beyond the scope of the present work.

\section{Conclusions\label{sec:conclusions}}

We have investigated the application of twisted boundary conditions to quenched
lattice QCD simulations of three-point correlation functions in order to access spatial 
components of hadronic momenta different from the integer multiples of $2\pi / L$. 
The vector and scalar form factors relevant to the $K \to \pi$ semileptonic decay have 
been evaluated by twisting in all possible ways one of the quark lines in the three-point 
functions. We have found that the momentum shift produced by the twisted boundary 
conditions does not introduce any additional noise and easily allows to determine 
within a few percent statistical accuracy the form factors at quite small values of 
the four-momentum transfer, which are not accessible when periodic boundary conditions 
are considered. We are confident that these findings are independent of the use 
of the quenched approximation, so that they will hold as well also in case of partially 
quenched and full QCD simulations.

We have studied the impact of twisted boundary conditions on the precision of the 
determination of the $K \to \pi$ form factors and their slopes at zero momentum 
transfer. We have found that: ~ i) when the precise lattice point for $f_0(q^2)$ at 
$q^2 = q_{\rm max}^2$ is not included in the analysis, the use of twisted BCs is crucial 
to allow a determination of the form factor at zero-momentum transfer at a few percent 
statistical level; ~ ii) when the very accurate value $f_0(q_{\rm max}^2)$ is included 
in the analysis, the impact of twisted BCs turns out to be marginal, while we remind it 
is expensive for the computational time. The latter result is due to the fact that: 
~ i) the precision of the lattice points obtained with twisted BCs is not comparable 
to the one that can be achieved for $f_0(q_{\rm max}^2)$; ~ ii) the use of twisted 
boundary conditions does not lead to a sufficient improvement of the precision with 
respect to the nearest points obtained with periodic boundary conditions.

We stress that there are cases of phenomenological interest where a lattice point at 
zero-momentum transfer is not accessible with periodic BCs, like e.g.~the case of the 
weak magnetism, the weak electricity, the induced scalar and pseudoscalar form 
factors entering hyperon semileptonic decays. 

\section*{Acknowledgements} The authors gratefully acknowledge V.~Lubicz, M.~Papinutto 
and G.~Villadoro for many useful discussions and comments.


\begin{thebibliography}{99}

\bibitem{Bedaque}
  P.~F.~Bedaque,
  Phys.\ Lett.\ B {\bf 593} (2004) 82
  [arXiv:nucl-th/0402051].
  
\bibitem{Tantalo}
  G.~M.~de Divitiis, R.~Petronzio and N.~Tantalo,
  Phys.\ Lett.\ B {\bf 595} (2004) 408
  [arXiv:hep-lat/0405002].

\bibitem{Kl3} 
  D.~Becirevic {\it et al.},
  Nucl.\ Phys.\ B {\bf 705} (2005) 339
  [arXiv:hep-ph/0403217] 
  and Nucl.\ Phys.\ Proc.\ Suppl.\  {\bf 140} (2005) 387.
  See also for hyperon semileptonic decays: D.~Becirevic {\it et al.},
  Eur.\ Phys.\ J.\ A {\bf 24S1} (2005) 69
  [arXiv:hep-lat/0411016].

\bibitem{Kl3_2df}
  C.~Dawson, T.~Izubuchi, T.~Kaneko, S.~Sasaki and A.~Soni,
  PoS {\bf LAT2005} (2005) 337
  [arXiv:hep-lat/0510018].
  N.~Tsutsui {\it et al.}  [JLQCD Collaboration],
  PoS {\bf LAT2005} (2005) 357
  [arXiv:hep-lat/0510068].

\bibitem{Kl3_MILC}
  M.~Okamoto  [Fermilab Lattice Collaboration],
  arXiv:hep-lat/0412044.

\bibitem{hyperons}
  D.~Guadagnoli, V.~Lubicz, M.~Papinutto and S.~Simula, preprint RM3-TH/06-7.

\bibitem{Sachrajda}
  C.~T.~Sachrajda and G.~Villadoro,
  Phys.\ Lett.\ B {\bf 609} (2005) 73
  [arXiv:hep-lat/0411033].
  J.~M.~Flynn, A.~Juttner and C.~T.~Sachrajda  [UKQCD Collaboration],
  Phys.\ Lett.\ B {\bf 632} (2006) 313
  [arXiv:hep-lat/0506016].

\bibitem{Bedaque_2}
  P.~F.~Bedaque and J.~W.~Chen,
  Phys.\ Lett.\ B {\bf 616} (2005) 208
  [arXiv:hep-lat/0412023].

\bibitem{S1} 
  G.~Martinelli and C.~T.~Sachrajda,
  Nucl.\ Phys.\ B {\bf 316} (1989) 355.

\bibitem{nucleon}
  M.~Gockeler {\it et al.} [QCDSF Collaboration],
  Phys.\ Rev.\ D {\bf 71} (2005) 034508
  [arXiv:hep-lat/0303019].

\bibitem{NEDM_1}
  E.~Shintani {\it et al.},
  Phys.\ Rev.\ D {\bf 72} (2005) 014504
  [arXiv:hep-lat/0505022].

\bibitem{NEDM_2}
  F.~Berruto, T.~Blum, K.~Orginos and A.~Soni,
  Phys.\ Rev.\ D {\bf 73} (2006) 054509
  [arXiv:hep-lat/0512004].

\bibitem{NEDM_3}
  D.~Guadagnoli, V.~Lubicz, G.~Martinelli and S.~Simula,
  JHEP {\bf 0304} (2003) 019
  [arXiv:hep-lat/0210044].
  D.~Guadagnoli and S.~Simula,
  Nucl.\ Phys.\ B {\bf 670} (2003) 264
  [arXiv:hep-lat/0307016].

\end{thebibliography}
\end{document}